\documentclass[amsmath,amssymb,reprint,aps,pra,floatfix,superscriptaddress]{revtex4-1}
\usepackage{bm}
\usepackage{graphicx}
\usepackage{braket}
\usepackage{epstopdf}
\usepackage{bm, bbm}
\usepackage{lineno}
\usepackage{natbib}

\begin{document}

\title{Projecting onto any two-photon polarization state using linear optics}

\author{G. S. Thekkadath}
\email{guillaume.thekkadath@physics.ox.ac.uk}
\affiliation{Department of Physics and Centre for Research in Photonics, University
of Ottawa, 25 Templeton Street, Ottawa, Ontario, K1N 6N5, Canada}
\affiliation{Clarendon Laboratory, University of Oxford, Parks Road, Oxford, OX1 3PU, UK}

\author{L. Giner}
\affiliation{Department of Physics and Centre for Research in Photonics, University
of Ottawa, 25 Templeton Street, Ottawa, Ontario, K1N 6N5, Canada}

\author{X. Ma}
\affiliation{Department of Physics and Centre for Research in Photonics, University
of Ottawa, 25 Templeton Street, Ottawa, Ontario, K1N 6N5, Canada}
\affiliation{School of Physics, Nankai University, 94 Weijin Road, Tianjin, 300071, People's Republic of China}

\author{J. Fl\'orez}
\affiliation{Department of Physics and Centre for Research in Photonics, University
of Ottawa, 25 Templeton Street, Ottawa, Ontario, K1N 6N5, Canada}

\author{J. S. Lundeen}
\affiliation{Department of Physics and Centre for Research in Photonics, University
of Ottawa, 25 Templeton Street, Ottawa, Ontario, K1N 6N5, Canada}

\begin{abstract}
Projectors are a simple but powerful tool for manipulating and probing quantum systems. For instance, projecting two-qubit systems onto maximally entangled states can enable quantum teleportation. While such projectors have been extensively studied, partially-entangling projectors have been largely overlooked, especially experimentally, despite their important role in quantum foundations and quantum information. Here, we propose a way to project two polarized photons onto any state with a single experimental setup. Our scheme does not require optical non-linearities or additional photons. Instead, the entangling operation is provided by Hong-Ou-Mandel interference and post-selection. The efficiency of the scheme is between 50\% and 100\%, depending on the projector. We perform an experimental demonstration and reconstruct the operator describing our measurement using detector tomography. Finally, we flip the usual role of measurement and state in Hardy's test by performing a partially-entangling projector on separable states. The results verify the entangling nature of our measurement with six standard deviations of confidence.
\end{abstract}

\maketitle

\section{Introduction}
In quantum physics, measurements are used for both controlling and probing quantum systems. The simplest measurement has two possible outcomes, 1 or 0, and is described by an operator $\bm{P}$ having a single eigenstate $\ket{\psi}$ with a non-zero eigenvalue, i.e. a projector $\bm{P}= \ket{\psi}\bra{\psi}$. Despite their simplicity, projectors are the archetypal measurement in many quantum information processing tasks such as secure key distribution~\cite{bb84}, state estimation~\cite{leonhardt1997measuring}, and testing Bell's inequalities~\cite{bell1964on,clauser1969proposed}. Usually, it is experimentally easy to project a single qubit onto any state. In the case of a photon's polarization, a combination of quarter-wave plate and polarizer can achieve any projector. However, quantum information processing (QIP) aims to leverage the resources that emerge in multi-photon systems, especially entanglement. Projecting multi-photon systems onto maximally entangled states can enable optical quantum computing and communication protocols, including quantum logic gates~\cite{hofmann2002quantum,obrien2003demonstration,kiesel2005linear,langford2005demonstration} and quantum teleportation~\cite{bouwmeester1997experimental}. 

Since entanglement is considered to be a resource in QIP, one might think maximally-entangling measurements are always more valuable than partially-entangling ones. However, there are many scenarios in which one may want to project onto a state with a tunable amount of entanglement. For example, in the usual teleportation protocol Alice and Bob ideally share a maximally entangled state. But if this state is imperfect and distillation~\cite{bennett1996concentrating} is not possible, then a partially-entangling projector optimizes Alice's success probability of transferring her qubit to Bob~\cite{li2000probabilistic,agrawal2002probablistic,modlawska2008nonmaximally,fortes2013improving}. Similarly, in the partial teleportation protocol~\cite{filip2004quantum,filip2004conditional,zhao2005experimental}, Alice performs a partially-entangling projector to transfer an imperfect copy of her qubit to Bob while still retaining an imperfect copy of her own. This second strategy achieves both asymmetric cloning~\cite{cerf2000pauli,firasek2001optical} and telecloning~\cite{murao1999quantum,murao2000quantum}.

Beyond QIP, partially-entangling measurements can be used to probe foundational issues in quantum physics. For example, the PBR theorem sheds light on the physical significance of the quantum state~\cite{pusey2012reality}. The theorem can be verified experimentally by performing a set of four projectors, two of which are partially-entangling. Similarly, the collective measurement proposed in Ref.~\cite{massar1995optimal} can be implemented via four partially-entangling projectors on pairs of identical qubits. This collective measurement is of fundamental interest as it optimally extracts information from a finite number of copies of a system~\cite{hou2017deterministic}.



All these applications motivate the need for a single measurement device capable of projecting a two-qubit system onto any desired state. In principle, this could be achieved using a CNOT gate combined with local operations on each qubit~\cite{nielsen2010quantum}. Although the CNOT gate has been realized experimentally with two-photon polarization states~\cite{obrien2003demonstration}, this approach is neither the simplest nor the most efficient. Other proposed schemes require complications such as ancilla photons~\cite{grudka2002projective,ahnert2006anypovm}. Here, we propose and experimentally demonstrate a straightforward scheme to measure the projector $\bm{P} = \ket{\psi}\bra{\psi}$ where
\begin{equation}
\label{eqn:general_twophoton}
\ket{\psi} = c_{HH}\ket{H_aH_b} + c_{HV}\ket{H_aV_b} + c_{VH}\ket{V_aH_b} + c_{VV}\ket{V_aV_b}
\end{equation}
is a general two-photon polarization state ($a$ and $b$ label two spatial modes, $H$ is horizontally polarized, and $V$ is vertically polarized).

\section{Theory}

\begin{figure}
\centering \includegraphics[width=0.4\textwidth]{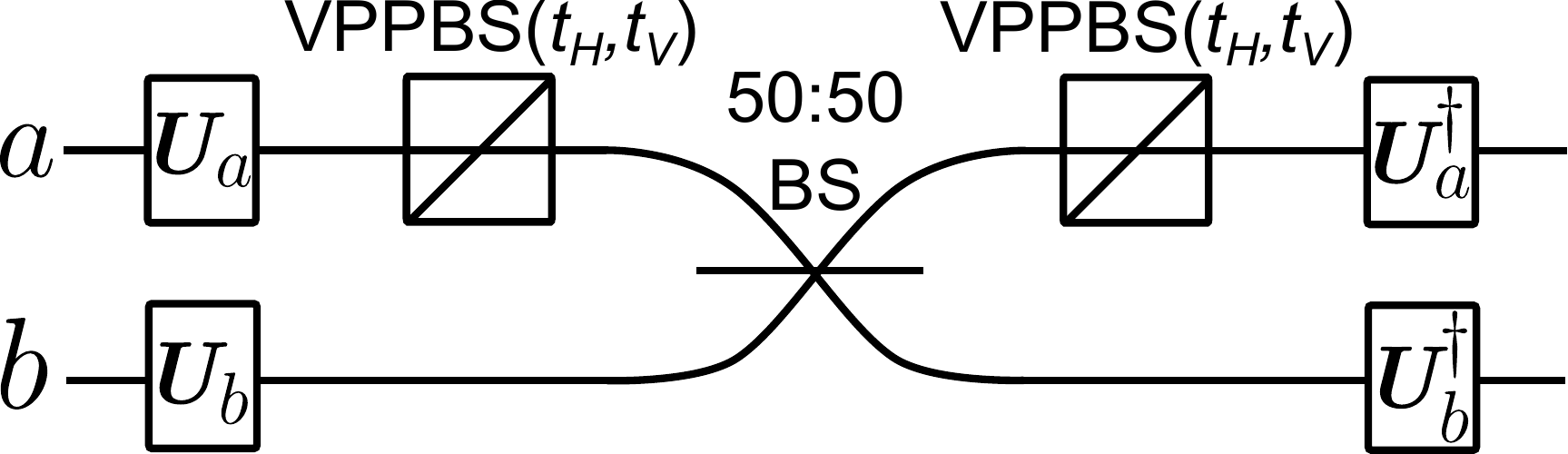} 
\caption{Schematic sketch of the scheme. The input system is two polarized photons in spatial modes $a$ and $b$. The projector works probabilistically by post-selecting on cases when each photon exits the 50:50 beam splitter (BS) into separate modes. Any projector can be measured by choosing the appropriate unitaries ($\bm{U}_a$ and $\bm{U}_b$), and transmission amplitudes ($t_H, t_V$) in each variable partially-polarizing BS (VPPBS).}
\label{fig:scheme} 
\end{figure}

The scheme is shown schematically in Fig.~\ref{fig:scheme}. It consists only of linear optical elements, such as wave plates and beam splitters, and does not require ancillas. In general, the state $\ket{\psi}$ that we wish to project onto may be entangled and thus our scheme needs an entangling operation. Due to poor photon-photon interactions, a deterministic entangling operation would require optical non-linearities or a complicated combination of ancillas and linear optical elements~\cite{knill2001scheme}. Taking inspiration from previous demonstrations of probabilistic quantum logic gates~\cite{hofmann2002quantum,obrien2003demonstration,kiesel2005linear,langford2005demonstration}, our entangling operation is provided by Hong-Ou-Mandel interference at a beam splitter (BS) along with post-selection. This is well studied in the context of Bell measurements where post-selection on two anti-bunched photons after a BS is used to project onto the maximally entangled anti-symmetric state~\cite{braunstein1995measurement,bouwmeester1997experimental}. By adding local unitaries (i.e. wave plates) before and after the BS, one can project onto any maximally entangled state. However, in order to be able to project onto partially entangled states, local unitaries do not suffice since these cannot decrease the entanglement of the projected state. Instead, we induce controllable polarization-dependent loss in one of the modes before and after the BS which imbalances the Hong-Ou-Mandel interference effect. This loss is achieved by a variable partially-polarizing BS, which we now describe in detail.

A polarizing BS completely separates horizontal ($H$) and vertical ($V$) light into separate spatial modes. A more general operation can be achieved by allowing the splitting ratios for $H$ and $V$ to be independent and tunable. Previous works used such partially-polarizing BSs in probabilistic quantum logic gates~\cite{hofmann2002quantum,obrien2003demonstration,kiesel2005linear,langford2005demonstration} and asymmetric cloning~\cite{zhao2005experimental,soubusta2008experimental}. However, in those experiments (with the exception of Ref.~\cite{soubusta2008experimental}), the BSs had fixed splitting ratios. Here we consider a device where the splitting ratios can be tuned, i.e. a variable partially-polarizing beam splitter (VPPBS)~\cite{florez2018variable}, so that any projector can be implemented with a single setup. A VPPBS acting on mode $a$ is described by the following transformation:
\begin{equation}
\begin{split}
\bm{a}_j^\dagger &\rightarrow t_j\bm{a}_j^\dagger + i(1-t^2_j)^{1/2}\bm{r}_j^\dagger \\
\end{split}
\label{eqn:vppbs_transf}
\end{equation}
where $\bm{a}^\dagger_j\ket{0} = \ket{j_a}$ is a creation operator, $r$ is the reflected mode, and $t_j\in [0,1]$ with $j=H,V$ are independently tunable real transmission amplitudes for $H$ and $V$ polarized light, respectively. By ignoring the reflected mode of the VPPBS, we can induce polarization-dependent loss in mode $a$. In the two-photon basis $\{ \ket{H_aH_b}, \ket{H_aV_b}, \ket{V_aH_b}, \ket{V_aV_b}\}$, the transformation $\bm{W}$ describing a VPPBS in mode $a$ and the identity operator in mode $b$ is:
\begin{equation}
\label{eqn:vppbs_matrix}
\bm{W} = 
\begin{pmatrix}
t_H & 0 & 0 & 0 \\
0 & t_H & 0 & 0 \\
0 & 0 & t_V & 0 \\
0 & 0 & 0 & t_V \\
\end{pmatrix}.
\end{equation}
We assumed that both $t_H$ and $t_V$ are real, but we note that there could be a non-zero relative phase $\delta$ between the two in a physical realization of the VPPBS.

After the VPPBS, both photons impinge onto different ports of a non-polarizing 50:50 BS. The photons are assumed to have the same spatial and spectral distributions, and arrive at the BS at the same time. As such, if the two photons are in a symmetric polarization state (i.e. their combined state is unchanged when the modes of both photons are swapped), they always leave the BS from the same port due to Hong-Ou-Mandel interference~\cite{braunstein1995measurement}. Hence, by post-selecting on cases where the photons exit the BS from different ports, we project onto the anti-symmetric singlet state $\ket{s}=(\ket{H_aV_b} - \ket{V_aH_b})/\sqrt{2}$:
\begin{equation}
\label{eqn:singlet_proj}
\ket{s}\bra{s} = \frac{1}{2} \begin{pmatrix}
0 & 0 & 0 & 0 \\
0 & 1 & -1 & 0 \\
0 & -1 & 1 & 0 \\
0 & 0 & 0 & 0 \\
\end{pmatrix},
\end{equation}
where the matrix is written in the same basis as Eq.~\ref{eqn:vppbs_matrix}.

Finally, a second VPPBS, with the same transmission amplitudes as the first one, is placed in mode $a$ after the 50:50 BS. The result of the entire sequence can be found by multiplying the three transformations in the correct order:
\begin{equation}
\label{eqn:projector_final}
\bm{W} \times \ket{s}\bra{s} \times \bm{W} = \eta \ket{\tilde{\psi}}\bra{\tilde{\psi}}
\end{equation}
where $\ket{\tilde{\psi}} = (\sqrt{1+\gamma}\ket{H_aV_b} - \sqrt{1-\gamma} \ket{V_aH_b}) / \sqrt{2}$, $\gamma = (t^2_H-t^2_V)/(t^2_H+t^2_V)$ is the VPPBS splitting ratio, and $\eta=(t^2_H+t^2_V)/2$ is an efficiency factor that we discuss later. We derive the same result using second quantization in Appendix~\ref{sec:sec_quant}. The quantity $|\gamma|$ sets the degree of entanglement (i.e. concurrence) $\mathcal{C}$ of $\ket{\tilde{\psi}}$ since $\mathcal{C}(\tilde{\psi}) = \sqrt{1-\gamma^2}$~\cite{wotters1998entanglement}. When $\gamma=0$ ($|\gamma|=1$), $\ket{\tilde{\psi}}$ is maximally entangled (separable). 

Any state $\ket{\psi}$ with a degree of entanglement set by $|\gamma|$ will have the same coefficients as $\ket{\tilde{\psi}}$ when written in its Schmidt basis. One can apply local unitaries $\bm{U}_a$ in mode $a$ and $\bm{U}_b$ in mode $b$ to transform $\ket{\psi}$ from its Schmidt basis to the $\{\ket{H},\ket{V}\}$ basis, that is: $\ket{\tilde{\psi}} = \bm{U}_a\bm{U}_b\ket{{\psi}}$. Thus, the remaining step to achieve the most general projector $\ket{\psi}\bra{\psi}$ (see Eq.~\ref{eqn:general_twophoton}) is to make the transformations $\bm{U}_a\bm{U}_b$ before the first VPPBS and $\bm{U}^\dagger_a\bm{U}^\dagger_b$ after the second VPPBS. These can be accomplished with a quarter-wave plate (QWP) and a half-wave plate (HWP) in each mode, as well as a birefringent element to control the phase $\delta$ between $t_H$ and $t_V$ (see Appendix~\ref{sec:schmidt})~\cite{wei2005synthesizing}. In that case, our scheme involves four wave plates angles (the angles for $\bm{U}^\dagger_a$ and $\bm{U}^\dagger_b$ are fixed by those used for $\bm{U}_a$ and $\bm{U}_b$, respectively), the phase $\delta$, and the VPPBS splitting ratio $\gamma$. All together, these comprise six degrees of freedom which is the same number as in a pure two-photon polarization state.

A successful projection is heralded by the presence of a photon in modes $a$ and $b$ at the output of the measurement device. This occurs with a probability given by the expectation value of the measurement operator in Eq.~\ref{eqn:projector_final} with respect to the state being measured $\ket{\phi_{in}}$, i.e. $\eta \left|\braket{\phi_{in}|\tilde{\psi}}\right|^2$. Since $\left|\braket{\phi_{in}|\tilde{\psi}}\right|^2$ is the ideal probability of a successful projection, we can treat $\eta$ as the efficiency of our scheme. The value of $\eta$ is unchanged by $\bm{U}_a\bm{U}_b$. In order to optimize $\eta$, the total VPPBS transmission $t_H^2 + t_V^2$ should be maximized for a given $\gamma\in [-1,1]$ to avoid unnecessary loss. If $\gamma\geq0$, this can be achieved by setting $t_H^2 =1$ and $t_V^2 = (1-\gamma)/(1+\gamma)$. Similarly, if $\gamma\leq0$, set $t_V^2 = 1$ and $t_H^2 = (1+\gamma)/(1-\gamma)$. Satisfying these conditions provides the optimal efficiency $\eta_{opt}$ which is given by:
\begin{equation}
\eta_{opt} = \frac{1}{1+|\gamma|} = \frac{1}{1+\sqrt{1-\mathcal{C}^2}}.
\end{equation}
Although $\eta_{opt}$ is independent of the input state being measured, it depends on the degree of entanglement $\mathcal{C}$ of the projector being performed. This is because post-selecting on anti-bunching after the 50:50 BS is an efficient way to project onto maximally entangled states ($\mathcal{C} = 1$, $\eta_{opt} = 1$) but not onto separable states ($\mathcal{C}=0$, $\eta_{opt} = 1/2$). While the latter can be achieved with unit efficiency using a simpler setup consisting of wave plates and polarizers, we stress the fact that our scheme is far more general.

The measurement device presented thus far is non-destructive since the two photons are in the state $\ket{\psi}$ whenever they exit the device from separate modes. More generally, quantum measurements do not necessarily leave the measured system in an eigenstate of the measurement operator. For example, a VPPBS in mode $a$ followed by a 50:50 BS and post-selection on anti-bunching implements the transformation $\bm{M}_{\tilde{\psi}} = \ket{s}\bra{s} \times \bm{W} = \eta^{1/2} \ket{s}\bra{\tilde{\psi}}$. Applied to some state $\ket{\phi_{in}}$, $\bm{M}_{\tilde{\psi}}$ projects the system onto the state $\ket{s}$ with a probability $\eta \left|\braket{\phi_{in}|\tilde{\psi}}\right|^2$. The measurement enacted by the transformation $\bm{M}_{\tilde{\psi}}$ can be described using the positive-operator valued measure (POVM) formalism, in which case the measurement operator is the POVM element $\bm{\Pi}=\bm{M}_{\tilde{\psi}}^\dagger \bm{M}_{\tilde{\psi}} = \eta \ket{\tilde{\psi}}\bra{\tilde{\psi}}$~\cite{wiseman2010quantum}. This operator looks the same as the projector in Eq.~\ref{eqn:projector_final} with the distinction that the measurement device leaves the system in the state $\ket{s}$ rather than $\ket{\tilde{\psi}}$. As before, this measurement can be generalized to an arbitrary state $\ket{\psi}$ by adding the appropriate local unitaries $\bm{U}_a\bm{U}_b$ before the VPPBS, i.e. $\bm{M}_{\psi} =  \eta^{1/2} \ket{s}\bra{\tilde{\psi}}\bm{U}_{a}\bm{U}_{b}$ where $\ket{\tilde{\psi}} = \bm{U}_{a}\bm{U}_{b}\ket{\psi}$. Thus, the projector $\bm{P} = \ket{\psi}\bra{\psi}$ can be achieved with a simpler experimental setup if it is not a requirement that the photons be in the state $\ket{\psi}$ after the measurement device.

There are many scenarios in which the post-measurement state of the system is not of importance. Perhaps the most obvious one is if two polarization-insensitive detectors are placed after the 50:50 BS, one in mode $a$ and the other in mode $b$. Then, the coincidence rate of both detectors is proportional to the expectation value $\braket{\phi_{in}|\bm{\Pi}|\phi_{in}}$. By varying $\ket{\phi_{in}}$ and keeping $\bm{\Pi}$ fixed, the measurement operator $\bm{\Pi}$ can be reconstructed using a technique known as detector tomography~\cite{lundeen2009tomography,feito2009measuring}. We demonstrate this idea experimentally in the next section.

\section{Experiment}
\label{sec:exp}

\begin{figure*}
\centering \includegraphics[width=1\textwidth]{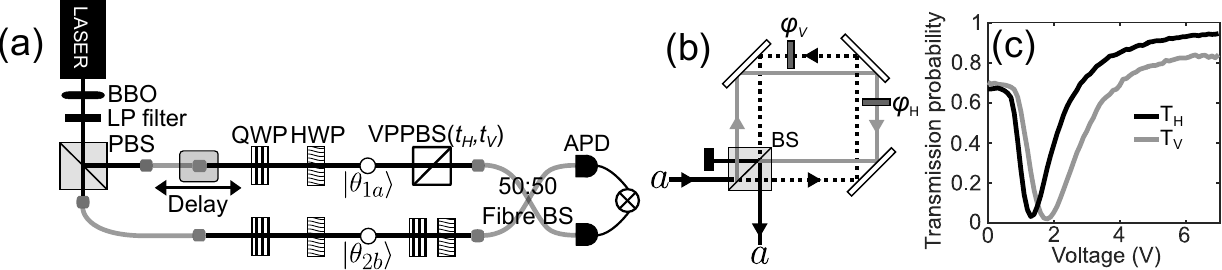} 
\caption{Experimental details. The experimental setup is shown in (a). The variable partially-polarizing beam splitter (VPPBS) is realized using the displaced Sagnac interferometer shown in (b). The phase between both paths in the interferometer is adjusted for both $H$ ($\varphi_H$) and $V$ ($\varphi_V$) polarizations independently using birefringent liquid crystals. In (c), we plot the VPPBS transmission probability $T_H$ (black line) and $T_V$ (grey line) measured when the input photon is $H$ and $V$ polarized, respectively, as a function of the voltage applied to the liquid crystal. As is common for such devices, the relation between voltage and retardance $\varphi_H$ and $\varphi_V$ is not linear. BBO: $\beta$-barium borate, LP: long pass, (P)BS: (polarizing) beam splitter, (Q/H)WP: (quarter/half) wave plate, APD: avalanche photodiode.}
\label{fig:fullsetup} 
\end{figure*}

The experimental setup is shown in Fig.~\ref{fig:fullsetup}a. A 404-nm-wavelength diode laser pumps a type-II $\beta$-barium borate crystal with 40 mW of power. Through spontaneous parametric down-conversion, pairs of 808-nm-wavelength photons with orthogonal polarization are generated collinearly with the pump laser. The latter is then blocked by a long pass filter. The photon pair splits at a PBS into modes $a$ and $b$. A pair of QWP and HWP in each mode is used to define the input state $\ket{\phi_{in}}$ (henceforth, $\bm{U}_a = \bm{U}_b =\mathbbm{1}$). Photons in path $a$ are sent into a VPPBS which we describe below. The exit of the VPPBS and path $b$ are then coupled into a single-mode-fibre non-polarizing 50:50 BS. A delay stage ensures that paths $a$ and $b$ have equal length such that the two photons can interfere. We pre-compensate for any polarization transformations in the fibre BS using an additional QWP and HWP pair in mode $b$. Finally, we measure the coincidence rate at the exit of the fibre	 BS using single-photon avalanche photodiodes (Excelitas SPCM-AQRH-24-FC).

The VPPBS (see Fig.~\ref{fig:fullsetup}b) consists of a displaced Sagnac interferometer which benefits from passive phase stability. The probability amplitude that the photon exits into mode $a$ depends on the relative phase between both paths in the interferometer. We adjust this relative phase for both $H$ and $V$ polarizations independently by introducing two phase shifters in the interferometer, one for each polarization. These phase shifters are birefringent liquid crystal cells with their optical axis aligned with either $H$ or $V$. Depending on the voltage applied to these liquid crystal cells, we can directly control the transmission probabilities $T_H$ and $T_V$, as shown in Fig.~\ref{fig:fullsetup}c. We observed a voltage-dependent phase $\delta$ between $t_H$ and $t_V$, i.e. $t_H = \sqrt{T_H}$ and $t_V = \sqrt{T_V}e^{i\delta}$. Due to limited interference visibility in the Sagnac interferometer ($\sim$ 93 \%), we can vary $T_H \in [0.03,0.95]$ and $T_V \in [0.02,0.84]$. This limits the range of projectors we can achieve experimentally, but does not affect their quality.

\subsection{Detector tomography}

\begin{figure}
\centering \includegraphics[width=0.35\textwidth]{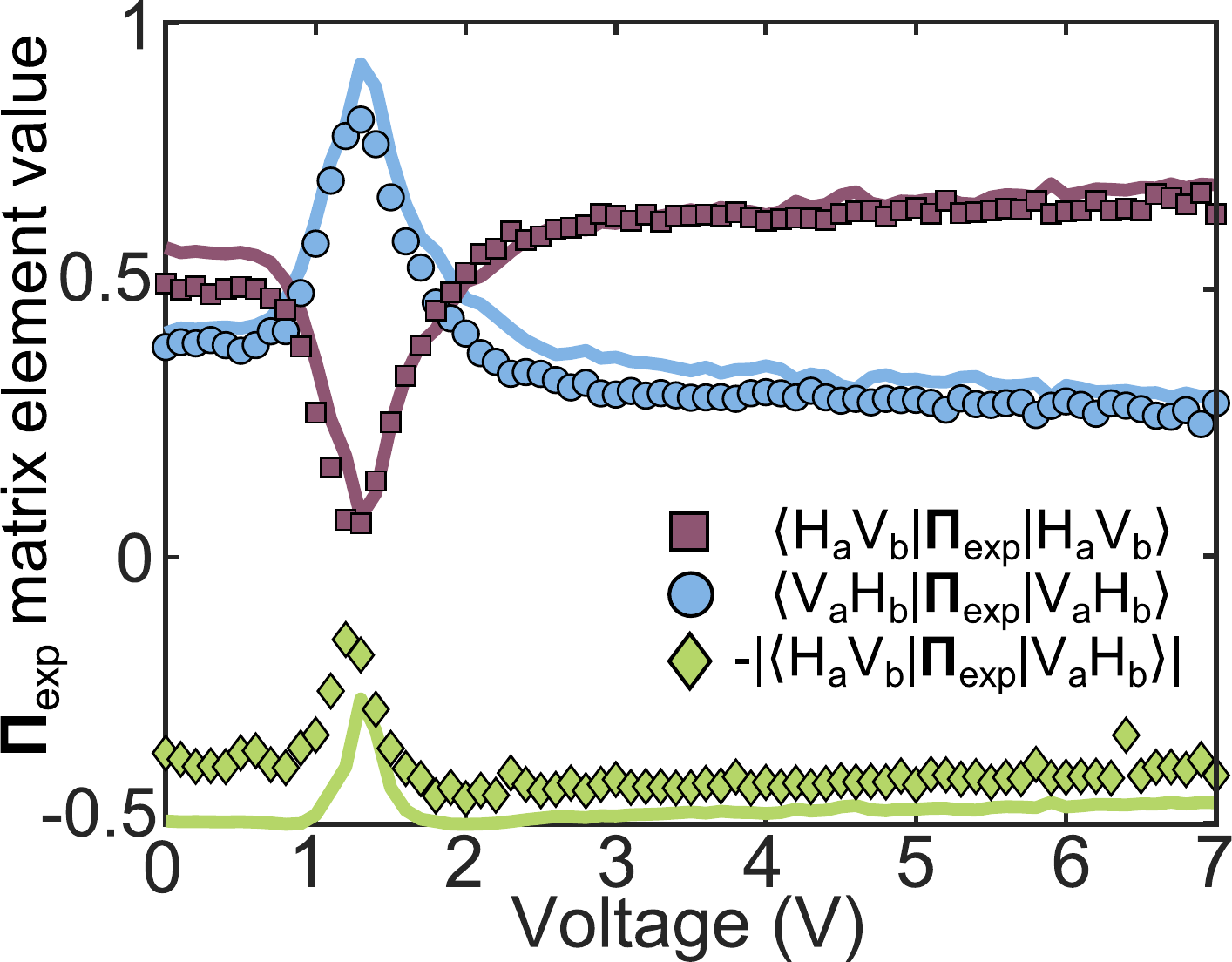} 
\caption{Detector tomography. We plot various matrix elements of the reconstructed $\bm{\Pi}_{exp}$ (markers) as a function of the voltage applied to the liquid crystal controlling $T_H$. The results can be compared to theory (bold lines).}
\label{fig:tomo_singlet} 
\end{figure}

Detector tomography~\cite{lundeen2009tomography,feito2009measuring} is the ideal tool to verify that our experimental setup performs the desired measurement. We treat our setup as an unknown measurement device and probe it by determining $\braket{\phi_{in}|\bm{\Pi}|\phi_{in}}$ for sixteen different input states, $\ket{\phi_{in}} \in \{\ket{H_aH_b}, \ket{V_aH_b}, \ket{D_aH_b}, \ket{R_aH_b},\ket{H_aV_b}, ... \}$, where $\ket{D} = (\ket{H} + \ket{V})/\sqrt{2}$ and $\ket{R} = (\ket{H} + i\ket{V})/\sqrt{2}$. The resulting counts are processed by a maximum-likelihood algorithm to reconstruct the closest matching positive and Hermitian operator $\bm{\Pi}_{exp}$. 

To demonstrate that various projectors can be achieved, we scan the VPPBS splitting ratio $\gamma$ by varying the voltage applied to the liquid crystal cell controlling $T_H$ and we fix $T_V = 0.458$. For each voltage step, we perform detector tomography and expect to reconstruct the operator $\bm{\Pi}= \eta \ket{\tilde{\psi}}\bra{\tilde{\psi}}$. However, both the efficiency $\eta$ and the projector $\ket{\tilde{\psi}}\bra{\tilde{\psi}}$ depend on $\gamma$ and thus the voltage. To distinguish between the two varying quantities, we normalize out $\eta$ from $\bm{\Pi}$, i.e. $\bm{\Pi}_{th}=\ket{\tilde{\psi}}\bra{\tilde{\psi}}$. Various matrix elements of the reconstructed $\bm{\Pi}_{exp}$ are shown in Fig.~\ref{fig:tomo_singlet}. Since the phase $\delta$ also varies with voltage, for clarity, we plot $-|\braket{H_aV_b|\bm{\Pi}_{exp}|V_aH_b}|$ so that the magnitude of this element can be compared to theory when $\delta=0$. The matrix elements not plotted are nearly zero ($\lesssim 0.035$), as expected. Also not shown is $\braket{V_aH_b|\bm{\Pi}_{exp}|H_aV_b}$ since it is identical to $\braket{H_aV_b|\bm{\Pi}_{exp}|V_aH_b}$. The bold lines are the expected values for the matrix elements of $\bm{\Pi}_{th}$ calculated using the measured transmission $T_H$ (shown in Fig.~\ref{fig:fullsetup}c) and fixing $T_V = 0.458$. We compute the fidelity $F=\mathrm{Tr}(\sqrt{\bm{\Pi}_{exp}}\bm{\Pi}_{th}\sqrt{\bm{\Pi}_{exp}})^{1/2}$ for each voltage step. Overall, we find an average of $F = 0.95$ with standard deviation $0.02$, which suggests that there is good agreement between experiment and theory.

As can be seen in Fig.~\ref{fig:tomo_singlet}, our scheme enables us to control the matrix elements of $\bm{\Pi}_{exp}$. In particular, we can control the degree of entanglement of the projected state. To quantify this, we compute the concurrence $\mathcal{C}$ of the reconstructed matrices $\bm{\Pi}_{exp}$ and find that we can vary $\mathcal{C} \in [0.13, 0.85]$. The lower bound of $\mathcal{C}$ is limited by the range over which we can vary the transmission probabilities $T_H$ and $T_V$, which in turn is limited by the interference visibility in our Sagnac interferometer ($\sim$ 93\%). This could be improved by using a different approach to implement the VPPBS, as discussed in Ref.~\cite{florez2018variable}. The upper bound of $\mathcal{C}$ is limited by the visibility of the quantum interference at the 50:50 BS ($\sim$ 90\%). In the next section, we describe the use of our setup to perform a partially-entangling projector that is of foundational importance for quantum mechanics.

\subsection{Hardy's test}
The violation of Bell's inequalities is the most convincing evidence that quantum mechanics cannot be described by a local hidden variable theory. Unfortunately, the derivation of the inequality is rather complicated and requires a number of involved logical steps before arriving at the final result~\cite{bell1964on,clauser1969proposed}. A more straightforward manifestation of the incompatibility of quantum mechanics with local realism is Hardy's test~\cite{hardy1993nonlocality,hardy1992quantum}. The arguments, which we outline below, are based on intuitive notions of joint probabilities that can be understood by the layman~\cite{kwiat2000mystery}.

Both Bell's and Hardy's tests require performing joint measurements on two entangled and space-like separated particles. Here we implement a ``reversed'' Hardy test by performing a partially-entangling projector on particles in separable states. That is, we flip the usual role of measurement and state in Bell's and Hardy's test since entanglement is used as a resource in the measurement rather than in the state preparation. Our measurement cannot, even in principle, be space-like separated. As such, although we measure the same joint probabilities as in Hardy's test, we cannot exclude the existence of local realism. Instead, a ``reversed'' test such as ours can certify the entangling nature of the measurement, which is necessary for protocols such as entanglement swapping~\cite{hensen2015loophole} or measurement-device-independent quantum key distribution~\cite{braunstein2012side,lo2012mdi,biham1996quantum,inamori2002security}.

\begin{figure}
\centering \includegraphics[width=0.4\textwidth]{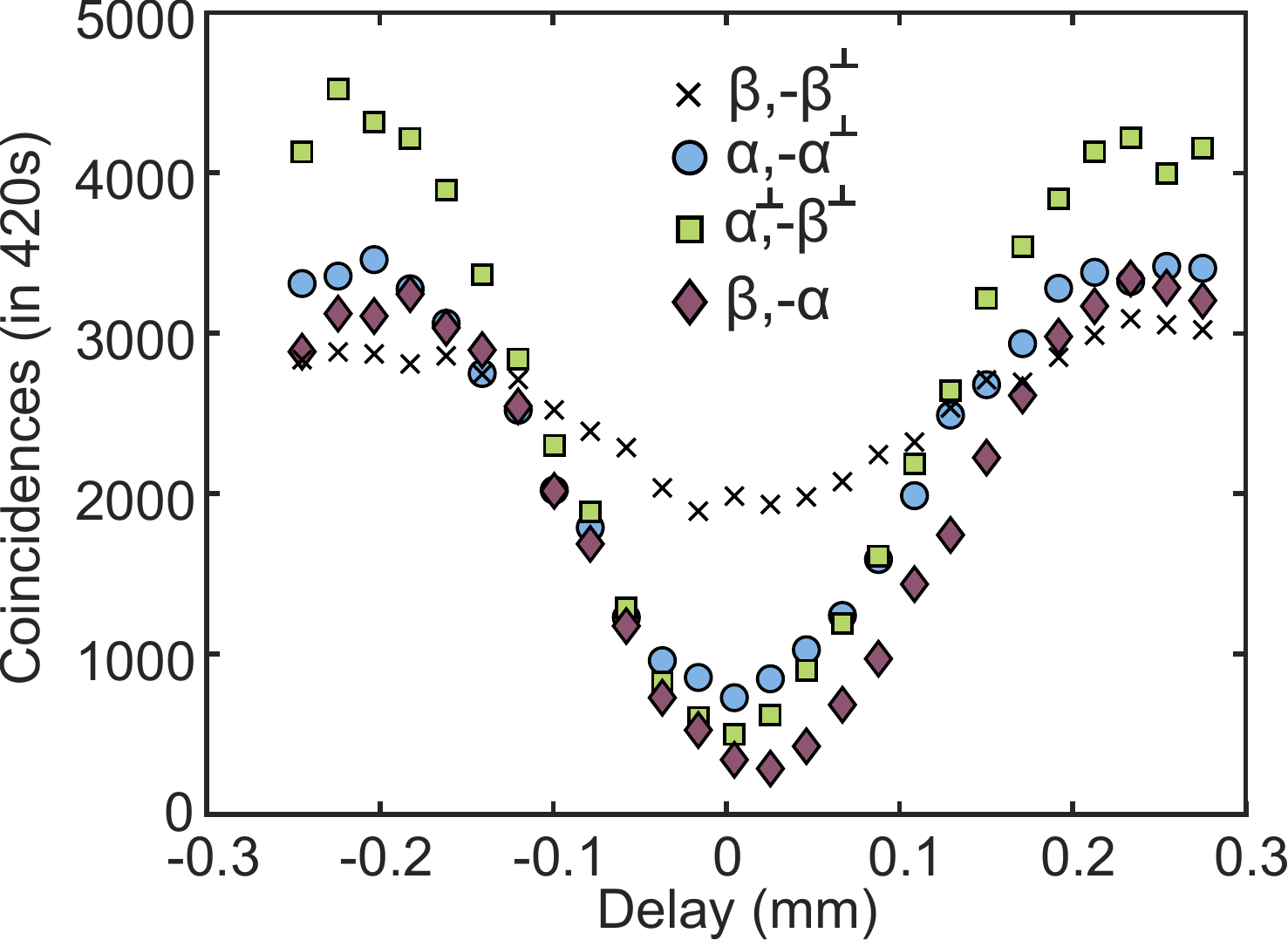} 
\caption{Hardy's test. We vary the delay between the photons in modes $a$ and $b$. At zero delay, Hong-Ou-Mandel interference can occur and we implement the measurement $\eta\ket{\tilde{\psi}}\bra{\tilde{\psi}}$ by post-selecting on coincidences. The coincidence rates at this point are shown in Table~I.}
\label{fig:hardy} 
\end{figure}

Suppose Alice sends the separable state $\ket{\phi_{in}}=\ket{\theta_{1a},\theta_{2b}}$ where $\ket{\theta_i} = \cos{\theta_i}\ket{H} + \sin{\theta_i}\ket{V} (i=1,2)$ into a black box concealing our measurement device and measures coincidences at its output. Since our device performs the projector $\eta \ket{\tilde{\psi}}\bra{\tilde{\psi}}$, the probability that Alice obtains coincidences is $P(\theta_1, \theta_2) = \eta \left|\braket{\theta_{1a}, \theta_{2b}|\tilde{\psi}}\right|^2 = \eta\left|\sqrt{1+\gamma}\cos{\theta_1}\sin{\theta_2}-\sqrt{1-\gamma}\sin{\theta_1}\cos{\theta_2}\right|^2/2$. Hardy showed that by choosing $\tan{\alpha} = [(1+|\gamma|)/(1-|\gamma|)]^{1/4}$ and $\tan{\beta}=-[(1+|\gamma|)/(1-|\gamma|)]^{3/4}$, then $P(\alpha, -\alpha^\perp) = P(\beta, -\alpha) = P(\alpha^\perp, -\beta^\perp) = 0$ (where $\alpha^\perp = \alpha + \pi/2)$~\cite{hardy1993nonlocality}. If Alice prepares these three states and measures no coincidences, she can infer that $P(\beta, -\beta^\perp) = 0$ for the following reasons. Alice realizes that the conditional probability of measuring coincidences when photon $b$ is $-\alpha^\perp$-polarized given photon $a$ is $\beta$-polarized is $P(\beta,-\alpha^\perp)/(P(\beta,-\alpha)+ P(\beta,-\alpha^\perp))=1$ (the denominator is the coincidence probability when photon $a$ is $\beta$-polarized). Similarly, the conditional probability of measuring coincidences when photon $a$ is $\alpha$-polarized given photon $b$ is $-\beta^\perp$-polarized is $P(\alpha,-\beta^\perp)/(P(\alpha,-\beta^\perp)+ P(\alpha^\perp,-\beta^\perp))=1$. Since Alice observes that $P(\alpha, -\alpha^\perp)=0$, in light of the previous statements, she concludes that she should measure $P(\beta, -\beta^\perp)=0$.

\begin{table}[h]
\centering
\begin{tabular}{c c}
Input state       					& Number of coincidences in 420 s\\
\hline
$\alpha, -\alpha^\perp$ 			& 727 \\
$\beta, -\alpha$           			& 340 \\
$\alpha^\perp, -\beta^\perp$        & 497 \\
$\beta, -\beta^\perp$   			& 1984\\
$\beta, -\alpha^\perp$   			& 1404 \\
$\alpha, -\beta^\perp$   			& 1391 \\

\end{tabular}
\label{table:hardy}
\caption{Measured concidences for Hardy's test.}
\end{table}

To her surprise, Alice in fact measures $P(\beta, -\beta^\perp) > 0$. By choosing $\gamma = 0.645$ ($\mathcal{C} = 0.764$), $P(\beta, -\beta^\perp)$ is maximized while the other three joint probabilities still vanish, meaning a partially-entangling projector is optimal for Hardy's test. In Fig.~\ref{fig:hardy}, we plot the coincidence rate as a function of the delay between the photons in modes $a$ and $b$. When the delay is zero, we implement the projector $\eta\ket{\tilde{\psi}}\bra{\tilde{\psi}}$ (we set $\delta = 0$ by tilting a wave plate about its axis). Due to experimental imperfections, we do not observe any vanishing coincidence rates (see Table I). There are two ways to deal with this. The first is to consider the inequality $N(\beta, -\beta^\perp)- N(\alpha, -\alpha^\perp) - N(\beta, -\alpha) - N(\alpha^\perp, -\beta^\perp) > 0$ where $N$ is the coincidence rate of each measurement~\cite{carlson2006quantum}. We find the left-hand-side to be $420 \pm 60$ and thus satisfy the inequality to within seven standard deviations. However, this inequality is susceptible to systematic errors since $N(\beta, -\beta^\perp)$ grows faster than the sum of the three other terms as $\gamma$ decreases and the input states are fixed. A second more convincing approach is to ask, given the measurement statistics in Table I, what can Alice infer about $N(\alpha, -\alpha^\perp)$ had she not measured that quantity~\cite{torgerson1995experimental}? She finds that $N(\beta, -\alpha^\perp)/(N(\beta, -\alpha) + N(\beta, -\alpha^\perp))=0.822 \pm 0.03$ instead of the ideal probability of one, as discussed earlier. Similarly, $N(\alpha, -\beta^\perp)/(N(\alpha, -\beta^\perp) + N(\alpha^\perp, -\beta^\perp))=0.737 \pm 0.03$. Then she would expect to measure a rate of $N(\alpha, -\alpha^\perp) = (0.822)(0.737)N(\beta, -\beta^\perp) = 1202 \pm 71$, which roughly six standard deviations larger than what she actually measures, $727$. Alice concludes that the measurement device concealed by the black box introduces correlations which cannot be described by her model. These non-classical correlations arise due to the entangling nature of the measurement.

\section{Conclusions}

In summary, we proposed a straightforward way to project two photons onto any polarization state. Our scheme has an efficiency of at least 50\% which far exceeds that of any scheme based on a probabilistic CNOT gate (11\%)~\cite{obrien2003demonstration}. We performed an experimental demonstration and reconstructed the operator describing our measurement using detector tomography. Finally, we flipped the usual role of measurement and state in Hardy's test and verified the entangling nature of our measurement.

While our proposal focuses on polarized photons, in principle it can be modified to work with qubits encoded in other internal degrees of freedom of a photon. Beam splitters can be used to project onto the anti-symmetric maximally entangled state for any variety of photonic qubits, e.g. time-bin~\cite{marcikic2003long}, orbital angular momentum~\cite{nagali2009optimal}, and temporal modes~\cite{mosley2008heralded}. One would also require a way to perform general unitary transformations (i.e. $\bm{U}_a\bm{U}_b$) and controllable loss (i.e. $\bm{W}$) on the degree of freedom of interest. For time-bin qubits this can be achieved by converting to a polarization encoding with fast switches~\cite{kupchak2017time}. Similarly, quantum pulse gates enable both rotations and controllable loss for photons in temporal modes~\cite{brecht2015photon}.

Partially-entangled states have already proven themselves to be a valuable resource in quantum physics~\cite{white1999nonmaxmimally}. For example, they are optimal to test Bell's theorem with imperfect detectors~\cite{eberhard1993background} and have been used in the recent landmark loophole-free tests~\cite{giustina2015significant,shalm2015strong}. Our scheme equips experimentalists with a simple tool to perform general two-qubit projectors. By extension, it also provides a tool to measure any POVM element on single qubits~\cite{nielsen2010quantum}. We hope that access to these new tools will stimulate work to find novel applications for partially-entangling measurements beyond the ones already mentioned in the introduction. For instance, the ability to project two qubits onto arbitrary states could be especially useful for state discrimination~\cite{barnett2009quantum} or quantum computing~\cite{howard2014contextuality,spekkens2008negativity}.

\begin{acknowledgments}
We thank Kevin Resch for discussions. This work was supported by the Canada Research Chairs (CRC) Program, the Canada First Research Excellence Fund (CFREF), and the Natural Sciences and Engineering Research Council (NSERC). G.S.T acknowledges support from the Oxford Basil Reeve Graduate Scholarship.
\end{acknowledgments}

\appendix
\onecolumngrid

\section{Derivation in second quantization}
\label{sec:sec_quant}
Here we describe the scheme in second quantization for additional clarity. We assume a general two-photon polarization state enters the measurement device:
\begin{equation}
\label{eqn:input_state}
\begin{split}
\ket{\phi_{in}} &= c_{HH}\ket{H_aH_b} + c_{HV}\ket{H_aV_b} + c_{VH}\ket{V_aH_b} + c_{VV}\ket{V_aV_b} \\
&=\left ( c_{HH}\bm{a}_H^\dagger\bm{b}_H^\dagger + c_{HV}\bm{a}_H^\dagger\bm{b}_V^\dagger + c_{VH}\bm{a}_V^\dagger\bm{b}_H^\dagger + c_{VV}\bm{a}_V^\dagger\bm{b}_V^\dagger \right ) \ket{0}_a\ket{0}_b,
\end{split}
\end{equation}
where $|c_{HH}|^2 + |c_{HV}|^2 + |c_{VH}|^2 + |c_{VV}|^2 = 1$. We evolve the creation operators in the Heisenberg picture using the input-output relations of the various components in the scheme.

In mode $a$, the photons undergo a VPPBS transformation:
\begin{equation}
\label{eqn:vppbs_in_out}
\bm{a}_j^\dagger \rightarrow t_j \bm{a}_j^\dagger + i(1-t^2_j)^{1/2}\bm{r}_{1j}^\dagger
\end{equation}
for $j=H,V$, and $r_{1}$ denotes the reflected (i.e. loss) mode of the VPPBS. Here, $t_H, t_V \in [0,1]$ are real transmission amplitudes of the VPPBS. Following the VPPBS, the photons impinge onto a 50:50 BS:
\begin{equation}
\begin{split}
\bm{a}_j^\dagger &\rightarrow (\bm{a}_j^\dagger + i\bm{b}_j^\dagger)/\sqrt{2} \\
\bm{b}_j^\dagger &\rightarrow (\bm{b}_j^\dagger + i\bm{a}_j^\dagger)/\sqrt{2}.
\end{split}
\end{equation}
Finally, the photons in mode $a$ undergo the VPPBS transformation with the same transmission amplitudes as in Eq.~\ref{eqn:vppbs_in_out}, i.e.
\begin{equation}
\bm{a}_j^\dagger \rightarrow t_j \bm{a}_j^\dagger + i(1-t^2_j)^{1/2}\bm{r}_{2j}^\dagger.
\end{equation}
Putting these transformations together yields a set of four input-output relations:
\begin{equation}
\label{eqn:in_out_arbproj}
\begin{split}
\bm{a}_H^\dagger &\rightarrow t_H\left(t_H\bm{a}_H^\dagger + i\bm{b}_H^\dagger + i(1-t_H^2)^{1/2}\bm{r}^\dagger_{2H}\right)/\sqrt{2} + i(1-t_H^2)^{1/2}\bm{r}^\dagger_{1H} \\
\bm{a}_V^\dagger &\rightarrow t_V\left(t_V\bm{a}_V^\dagger + i\bm{b}_V^\dagger + i(1-t_V^2)^{1/2}\bm{r}^\dagger_{2V}\right)/\sqrt{2} + i(1-t_V^2)^{1/2}\bm{r}^\dagger_{1V} \\
\bm{b}_H^\dagger &\rightarrow \left(\bm{b}_H^\dagger + it_H\bm{a}_H^\dagger - (1-t_H^2)^{1/2}\bm{r}^\dagger_{2H}\right)/\sqrt{2} \\
\bm{b}_V^\dagger &\rightarrow \left(\bm{b}_V^\dagger + it_V\bm{a}_V^\dagger - (1-t_V^2)^{1/2}\bm{r}^\dagger_{2V}\right)/\sqrt{2}.
\end{split}
\end{equation}
With these relations, we can find how the input state in Eq.~\ref{eqn:input_state} is transformed after the three components. Our scheme requires post-selecting on cases where the transformation leads to a photon in mode $a$ and mode $b$. Thus, when plugging the relations in Eq.~\ref{eqn:in_out_arbproj} into Eq.~\ref{eqn:input_state}, we only keep the terms $\bm{a}^\dagger_H\bm{b}^\dagger_H, \bm{a}_V^\dagger\bm{b}_V^\dagger, \bm{a}^\dagger_H\bm{b}^\dagger_V, \bm{a}_V^\dagger\bm{b}_H^\dagger$. With this post-selection, the resulting transformation is:
\begin{equation}
\begin{split}
\ket{\phi_{in}} \rightarrow \ket{\phi_{out}} &= \frac{1}{2} \left ( c_{HV} (t_H^2 \bm{a}_H^\dagger\bm{b}_V^\dagger - t_Ht_V \bm{a}_V^\dagger\bm{b}_H) + c_{VH} (t_V^2 \bm{a}_V^\dagger\bm{b}_H^\dagger - t_Ht_V \bm{a}_H^\dagger\bm{b}_V)  \right )\ket{0}_a\ket{0}_b \\
&=\frac{1}{2} \left( c_{HV} t_H^2 \ket{H_aV_b} - c_{HV}t_Ht_V \ket{V_aH_b} + c_{VH}t_V^2 \ket{V_aH_b} - c_{VH}t_Ht_V \ket{H_aV_b} \right) \\
&= \frac{1}{2}
\begin{pmatrix}
0 & 0 & 0 & 0 \\
0 & t^2_H & -t_Ht_V & 0 \\
0 & -t_Ht_V & t^2_V & 0 \\
0 & 0 & 0 & 0
\end{pmatrix}
\ket{\phi_{in}},
\end{split}
\end{equation}
where the matrix is written in the same notation as the main text. The matrix can be written as the outer product $\eta \ket{\tilde{\psi}}\bra{\tilde{\psi}}$ where $\ket{\tilde{\psi}} = (\sqrt{1+\gamma}\ket{H_aV_b} - \sqrt{1-\gamma} \ket{V_aH_b}) / \sqrt{2}$ is a normalized state, $\eta = (t_H^2 + t_V^2)/2$, and $\gamma=(t_H^2 - t_V^2)/(t_H^2 + t_V^2)$. Thus, we arrive at the same result as Eq.~\ref{eqn:projector_final}.

\section{Determining unitaries via a Schmidt decomposition}
\label{sec:schmidt}
The state $\ket{\psi} = c_{HH}\ket{H_aH_b} + c_{HV}\ket{H_aV_b} + c_{VH}\ket{V_aH_b} + c_{VV}\ket{V_aV_b}$ can always be written in the following form via a Schmidt decomposition:
\begin{equation}
\ket{\psi} = \lambda_1\ket{\zeta_a\theta_b} - \lambda_2\ket{\zeta^\perp_a\theta^\perp_b} 
\end{equation}
where $\lambda_1^2 + \lambda_2^2 = 1$ (both real numbers), and $\left|\braket{\zeta|\zeta^\perp}\right|=\left|\braket{\theta|\theta^\perp}\right|=0$. The unitaries $\bm{U}_a$ and $\bm{U}_b$ are found by solving the equations $\bm{U}_a\ket{\zeta} = \ket{H}$ and $\bm{U}_b\ket{\theta} = \ket{V}$, respectively. Since the right-hand-side of both equations is a linearly polarized state, each unitary can be accomplished with a quarter-wave plate and half-wave plate~\cite{sit2017general}. The final step to relate $\ket{\tilde{\psi}}$ to $\ket{\psi}$ is to set the transmission coefficients of the VPPBS to $t_H/(t^2_H + t^2_V)^{1/2} = \lambda_1$ and $t_V/(t^2_H + t^2_V)^{1/2} = \lambda_2$. Since we assumed that $\lambda_1$ and $\lambda_2$ are real, any relative phase $\delta$ between $t_H$ and $t_V$ should be compensated for using a birefringent element.



\bibliographystyle{apsrev4-1}
\bibliography{refs}

\end{document}